# *Large Area X-ray Proportional Counter (LAXPC) Instrument on AstroSat*

J. S. Yadav[1], P. C. Agrawal [2], H. M. Antia [1], R. K. Manchanda [3], B. Paul[4], Ranjeev Misra [5]

[1] Tata Institute of Fundamental Research, Homi Bhabha Road, Mumbai 400005, India
[2] UM-DAE Center of Excellence for Basic Sciences, University of Mumbai, Kalina, Mumbai-400098
[3] University of Mumbai, Kalina, Mumbai-400098, India
[4] Dept. of Astronomy & Astrophysics, Raman Research Institute, Bengaluru-560080 India
[5] Inter-University Center for Astronomy and Astrophysics, Ganeshkhind, Pune 411007, India

# *Abstract*

*Large Area X-ray Proportional Counter (LAXPC) is one of the major AstroSat payloads. LAXPC instrument will provide high time resolution X-ray observations in 3-80 keV energy band with moderate energy resolution. A cluster of three co-aligned identical LAXPC detectors is used in AstroSat to provide large collection area of more than 6000 $cm^2$. The large detection volume (15 cm depth) filled with xenon gas at ~ 2 atmosphere pressure, results in detection efficiency greater than 50%, above 30 keV. With its broad energy range and fine time resolution (10 microsecond), LAXPC instrument is well suited for timing and spectral studies of a wide variety of known and transient X-ray sources in the sky. We have done extensive calibration of all LAXPC detectors using radioactive sources as well as GEANT4 simulation of LAXPC detectors. We describe in brief some of the results obtained during the payload verification phase along with LXAPC capabilities.*

Astrosat is India`s first space astronomy observatory for simultaneous multi-wavelength studies with five science payloads. There are four X-ray instruments on-board AstroSat which cover wide energy band. These X-ray instruments are: (i) Large Area X-ray Proportional Counters (LAXPC) instrument covering 3-80 keV region, (ii) a Cadmium-Zinc-Telluride Imager (CZTI) array covering 20-100 keV, (iii) a Soft X-ray Imaging Telescope (SXT) covering 0.3 - 8 keV and (iv) a Scanning Sky Monitor (SSM) with energy range of 2 - 10 keV [1]. The main objective of LAXPC instrument is to study X-ray timing and wide band spectral properties of stellar and galactic systems containing compact objects.

The LAXPC instrument uses a *cluster of three co-aligned identical LAXPC detectors to achieve large area of collection in excess of 6000 $cm^2$. The deep detection volume (15 cm depth) filled with xenon gas at ~ 2 atmosphere*

*pressure, results in detection efficiency greater than 50%, above 30 keV. The LAXPC instrument consists of three identical units, each with its own independent front-end electronics, HV supply, and signal processing electronics. The system based time generator (STBG), is common for all the three LAXPCs to provide time stamp with accuracy of 10 micro-sec for all the accepted events. Each LAXPC detector consists of 60 anode cells of 3 cm x 3 cm cross-section and length of 100 cm, arranged in 5 layers providing a 15 cm deep X-ray detection volume. The Veto layer consisting of 1.5 cm x 1.5 cm x 100 cms anode cells, is divided in three parts (left side, right side and bottom) providing three Veto Layer outputs [2, 3]. The data from all the LAXPCs are independently acquired preserving the identity of each unit.*

The LAXPC is a large and complex X-ray instrument (10 flight packages with about 150 electronic cards including spare model) which has been designed and developed indigenously at TIFR, Mumbai. We have done extensive calibration of LAXPC detectors in the laboratory with radioactive sources as well as using GEANT4 simulation. AstroSat was launched successfully on 28[th] September, 2015. The STBG unit , processing electronics and low voltage detector electronics were switched on during 2[nd] and 3[rd] day after the launch. The LAXPC payload became fully functional on 19 th October, 2015 when high voltage (HV) of all three LAXPC detectors was switched on. Onboard LAXPC detector gas purification system was operated during 20-22 October & 23-24 November, 2015. LAXPC observation of CAS A supernova remnant is shown in Figure 1 which suggests around 20% energy resolution at 6.4 keV Iron line [2]. At higher energy (>20keV) , the energy resolution is found to be in the range of 10-14% [5].

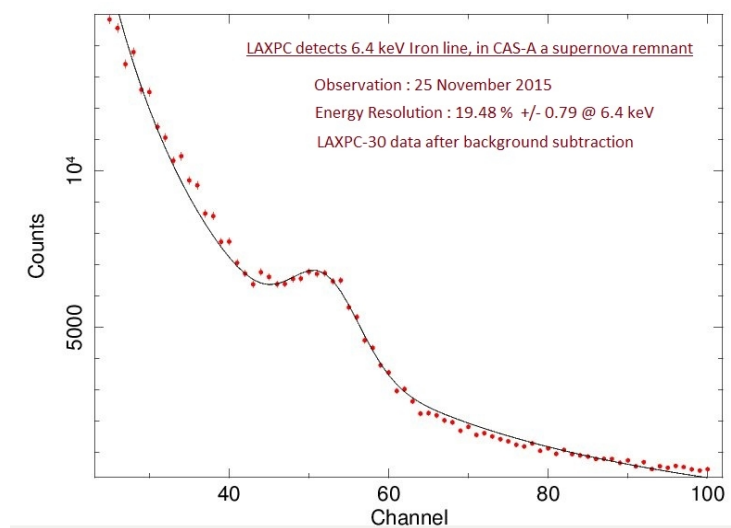

*Figure 1: CAS–A (Supernova remnant) observation of LAXPC30 which shows ~20% energy resolution at 6.4 keV.*

During the performance verification phase, AstroSat observed the black hole system GRS 1915+105 during 5-7 March, 2016 which was in the steep power law (SPL) state. To test the timing characteristics of the detector, event mode data which gives the energy of each photon and its arrival time with a time-resolution of 10 microsec is used to calculate the power density spectrum (PDS) up to the Nyquist frequency of 50 kHz which is shown in figure 2. The resulting PDS does not show any instrumental effect other than peaks beyond 10 kHz due to the dead-time of the detector (42.3 microsec). The green line show the expected peak power of a quasi-periodic oscillation (QPO) with quality factor Q=4 and rms of 5%. This LAXPC instrument can detect QPOs of such a strength easily till 3000 Hz. Energy dependent power spectra reveal a strong low-frequency (2–8 Hz) QPO and its harmonic along with broadband noise. [4] At the QPO frequencies, the time-lag as a function of energy has a non-monotonic behavior such that the lags decrease with energy till about 15–20 keV and then increase for higher energies.

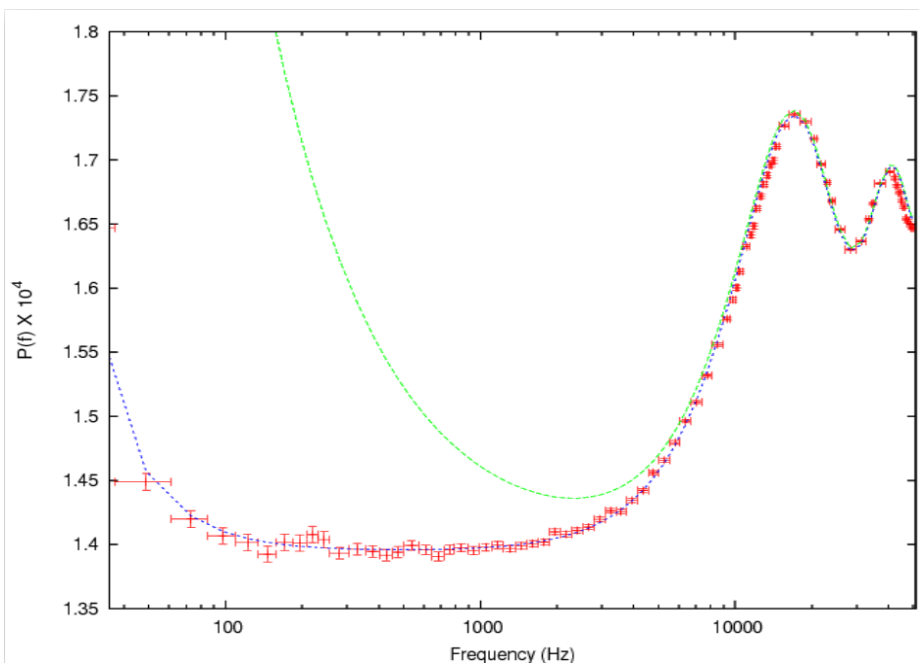

*Figure 2. High-frequency rebinned PDS of GRS 1915+105 in the SPL class observed during 5-7 March 2016. The power spectrum matches well with the predicted Poisson noise level with a dead time of ∆t = 42.3 microseconds and a low-frequency power-law component. The green line shows the expected peak power of a QPO with quality factor Q=4 and rms of 5% [4].*

LAXPC detectors have been extensively studied for effective area, detector response and background from GEANT4 simulation and compared to match with the laboratory calibration [5]. AstroSat observed another black hole system Cygnus X-1 in the low hard state which shows prominent thermal Comptonizaton component. The power spectrum can be characterized by two broad lorentzian functions centered at ~ 0.4 and ~ 3 Hz [6]. During the PV phase, the neutron star X-ray binary 4U 1728-34 was observed with AstroSat/LAXPC on 8th March 2016. We have detected typical Type-1 thermonuclear bursts in this source. Dynamical power spectrum of the data in the 3-20 keV band, reveals the presence of a kHz QPO whose frequency drifted from ~ 815 Hz at the beginning of the observation to about 850 Hz just before the burst [7]. This kHz QPO is also detected in the 10-20 keV band, which was not detected in earlier RXTE/PCA observations of this source.

Among the recent X-ray space missions, NASA`s Rossi X-ray Timing Experiment/Proportional Counter Array (RXTE/PCA) has been one of the most successful X-ray space missions [12]. The LAXPC instrument provides several advantages as compared to the RXTE/PCA. The effective area of the LAXPC at energies greater than 30 keV is significantly larger than the PCA [4, 5]. The event mode data allows for an energy dependent analysis of any choice of energy and time bins. Simultaneous observations with other instruments on board AstroSat, especially the Soft X-ray Telescope (SXT), can provide critical spectral coverage below 3 keV. LAXPC results discussed above demonstrate that LAXPC has advantage in the study of QPOs and their associated characteristics over RXTE/PCA specially in case of High Frequency (HF) QPOs. With largest effective area in the hard X-rays for any X-ray astronomy instrument ever flown, the LAXPCs will be particularly suited to exploit the timing and broad band spectroscopy in the 3-80 keV band. Here we highlight some key scientific topics that can be probed particularly well with the LAXPCs.

Accretion disk and relativistic jets are integral parts of black holes on all mass scales. Stellar mass black hole X-ray binaries often referred as microquasars provide favorable observational conditions like fast variability and a clear view of inner accretion disk where most of high energy processes are supposed to take place. Many of these are transient in nature. One of the key questions is how does energy flow during an outburst in a transient system which results in different X-ray states at different stages of an outburst and how relativistic transient radio jets are produced. The fast transitions between the low hard state (LH) and the high soft state (HS) has been seen so far only in two sources; GRS 1915+105 and IGR J17091-3624 [8, 9]. These fast transitions are attributed to thermal instability in the accretion disk but what causes this instability is not clear. LAXPC instrument can help in understanding evolution of outbursts in microquasars, evolution of different X-ray states and its connection with transient radio jets. Black hole systems are associated with intense gravitational fields and X-ray observations are a probe to test the theories of gravity. The mass and the spin of the black-hole are factors which determine the nature of the gravitational field. The black-hole spin can be measured independently using spectroscopy by modeling the X–ray spectrum and the relativistic broadening of the iron line. The Quasi Periodic Oscillations (QPOs) observed in the power density spectra of such systems arising from Keplerian period of the inner most stable orbit, can also be used to estimate the black-hole spin.

The accreting X-ray pulsars are neutron stars with very high magnetic field that accrete matter from a companion star and produce X-rays with a light house effect. These are the X–ray sources with a large fraction of their X-ray flux emitted in the hard X-ray band (greater than 10 keV) and are particularly suited for observations with the LAXPCs. The X-rays, having been produced in the polar regions of an assumed dipolar magnet, pass through the magnetosphere of the neutron star that is teeming with electrons and ions. Resonance scattering of the X-ray photons from these electrons in a strong magnetic field produce absorption feature in the X-ray spectrum, called cyclotron lines. It is a tell-tale signature of the extremely high magnetic field of the neutron stars, a few times $10^{12}$ gauss or more than a million times stronger than the strongest man made magnetic field. The LAXPCs, with very large effective area will enable study of the variation of the cyclotron line at different viewing angles and thus provide important clues regarding the structure of the magnetic field of the neutron stars, hitherto assumed to be dipole.

During the first week of operation when the LAXPC instrument became fully functional for the first time in orbit, the Be X-ray Binary 4U 0115+63 underwent a huge Type II outburst reaching a peak flux value of about 0.8 Crab. This binary has a neutron star as the X-ray source from which 3.61 sec pulsations have been detected. The LAXPC instrument made pointed mode observations of this binary on 2015, October 24 when the source intensity was near its peak. The light curves of three LAXPCs show strong intensity oscillations with a period of ~ 1000 sec ( or 1 mHz QPO) as can be seen in Fig 3. The oscillations arise from accretion disk instability [11].

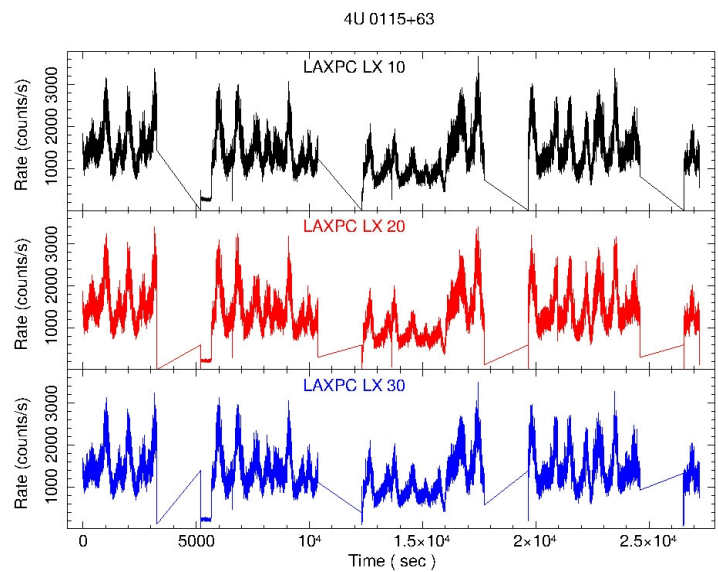

*Figure 3: AstroSat/LAXPC instrument observed 4U 0115-63 during an outburst on 5[th] day of its operation (24[th] October, 2015). The X-ray light curves are shown here as observed by three LAXPC detectors in 3-80 keV energy band.*

Some low magnetic field neutron stars, when accreting at a modest rate, show spectacular thermonuclear bursts. X-ray temperature and flux variation during these bursts is the most commonly used method for measurement of the radius of a neutron star. Such studies have so far been carried out in limited energy band and with assumption of simple black body type emission during the bursts. However, the non-burst emission from such neutron stars are known to suffer from reprocessing in the surrounding medium. With a wide energy band of the LAXPCs and large photon collection area, the thermonuclear burst spectroscopy will be performed to investigate the signatures of reprocessing, thus putting the prevailing method of neutron star radius measurement to critical examination. LAXPC has already observed thermonuclear bursts in two neutron star X-ray binaries; 4U 1728-34 and 4U 1636-536 [7, 10]. Along with the Ultra-Violet telescope onboard AstroSat, it will also be possible to investigate the reprocessing of the thermonuclear bursts in multiple optical and visible bands, a task that has been achieved only once in the past.

A class of binary X-ray sources, named Supergiant Fast X-ray Transients (SFXTs) have puzzled astronomers for over a decade. In spite of having binary components identical to a well known class of sources, the high mass X-ray binaries, they are less luminous by 2-4 orders of magnitude and show occasional X-ray flares that last a few minutes to a few hours. Different models, accretion from dense clumps in the winds of the companion star, gated accretion onto neutron stars with strong magnetic field, etc. have been proposed to explain the SFXT phenomena. LAXPC, along with the SXT of AstroSat will allow deep search for pulsation and/or cyclotron line from these objects and thus provide definitive information about the compact objects in them. If found, X-ray pulsations will be very useful to measure the binary parameters of the SFXTs with up to a few day orbital period.

Accreting neutron stars are potential sources of continuous Gravitational Wave (GW) emission. An accretion mound on a neutron star with spin frequency of several hundred Hertz can be a potential candidate for detection of continuous GW. However, to carry out search for GW in a long observation of a few months or years, spin and orbital parameters of the neutron star and their changes must be known over the entire period. Dedicated LAXPC observations could be utilized to search for pulsation of the neutron star and then determine the orbital parameters in some of the brightest accreting neutron stars in low mass X-ray binary systems that will aid search for continuous GW from them.

We acknowledge the strong support from Indian Space Research Organization (ISRO) in various aspect of instrument building, testing, software development and mission operation during payload verification phase. We acknowledge support of the scientific and technical staff of the LAXPC instrument team as well as staff of the TIFR Workshop in the development and testing of the LAXPC instrument.